\documentclass[9pt,twocolumn,twoside]{pnas-new}
\templatetype{pnasresearcharticle} % Choose template 
\usepackage{graphicx, xcolor, multirow}
\usepackage{siunitx, amssymb, amsmath}

\title{Molecular origin of the reduced phase stability and faster formation/dissociation kinetics in confined methane hydrate}

\author[a]{Dongliang Jin}
\author[a,1]{Benoit Coasne} 

\affil[a]{Univ. Grenoble Alpes, CNRS, LIPhy, 38000 Grenoble, France}
\leadauthor{Jin and Coasne} 

\authorcontributions{Both authors designed the work. D. Jin carried out the molecular simulations while both authors analyzed the data and developed the model. Both authors contributed to the writing of the manuscript.}
\authordeclaration{The authors declare no competing interests.}
\correspondingauthor{\textsuperscript{1}To whom correspondence should be addressed. E-mail: benoit.coasne\@univ-grenoble-alpes.fr}


\begin{abstract}
The microscopic mechanisms involved in the formation/dissociation of methane hydrate confined at the nanometer scale are unraveled using advanced molecular modeling techniques combined with a mesoscale thermodynamic approach. By means of atom-scale simulations probing coexistence upon confinement and free energy calculations, phase stability of confined methane hydrate is shown to be restricted to a narrower temperature and pressure domain than its bulk counterpart. The melting point depression at a given pressure, which is consistent with available experimental data, is shown to be quantitatively described using the Gibbs--Thomson formalism if used with accurate estimates for the pore/liquid and pore/hydrate surface tensions. The metastability barrier upon hydrate formation and dissociation is found to decrease upon confinement, therefore providing a molecular scale picture for the faster kinetics observed in experiments on confined gas hydrates. By considering different formation mechanisms -- bulk homogeneous nucleation,  external surface nucleation, and confined nucleation within the porosity -- we identify a crossover in the nucleation process; the critical nucleus formed in the pore corresponds either to a hemispherical cap or a bridge nucleus depending on temperature, contact angle, and pore size. Using the classical nucleation theory, for both mechanisms, the typical induction time is shown to scale with the pore surface to volume ratio and, hence, the reciprocal pore size. These findings for the critical nucleus and nucleation rate associated to such complex transitions provide a mean to rationalize and predict methane hydrate formation in any porous media from simple thermodynamic data.
\end{abstract}

\dates{This manuscript was compiled on \today}
\doi{\url{www.pnas.org/cgi/doi/10.1073/pnas.XXXXXXXXXX}}

\begin{document}

\maketitle
\thispagestyle{firststyle}
\ifthenelse{\boolean{shortarticle}}{\ifthenelse{\boolean{singlecolumn}}{\abscontentformatted}{\abscontent}}{}

% If your first paragraph (i.e. with the \dropcap) contains a list environment (quote, quotation, theorem, definition, enumerate, itemize...), the line after the list may have some extra indentation. If this is the case, add \parshape=0 to the end of the list environment.
\dropcap{M}ethane hydrate is a non-stoichiometric crystalline phase in which water molecules form hydrogen-bonded cages that entrap methane molecules~\cite{sloan_fundamental_2003, koh_state_2012}. Under typical terrestrial and marine conditions, methane hydrate forms according to a structure known as sI where 46 water molecules form two small pentagonal dodecahedral cages and six tetracaidecahedral cages so that 8 methane molecules can be encapsulated at most~\cite{sloan_clathrate_2007}. While the large methane hydrate resources available on Earth are still regarded as an important fossil energy source~\cite{macdonald_thefuture_1990}, their abundant presence in locations such as seafloor or permafrost is also a threat to the environment as methane is one of the worst greenhouse gases~\cite{serov_postglacial_2017, phrampus_recent_2012}. For instance, in the context of increasing concerns about climate change, even weak perturbations such as those induced by a small temperature raise could trigger the dissociation of methane hydrate and release of large amounts of methane into the atmosphere~\cite{wood_decreased_2002}. Gas hydrates including methane hydrate are also thought to be a role player in the geochemistry of planets, comets, etc. as typical temperature $T$ and pressure $P$ in many of these systems should promote their formation whenever water and gases are present~\cite{formisano_detection_2004}. Finally, methane hydrate is also of particular relevance to energy and environmental science with applications for energy storage and carbon capture~\cite{lee_tuning_2005, patt_grand_2018, english_mechanisms_2009, petuya_selective_2018, pohlman_enhanced_2017}.

\begin{SCfigure*}[\sidecaptionrelwidth][t]
        \centering
        \includegraphics[width=11.4cm]{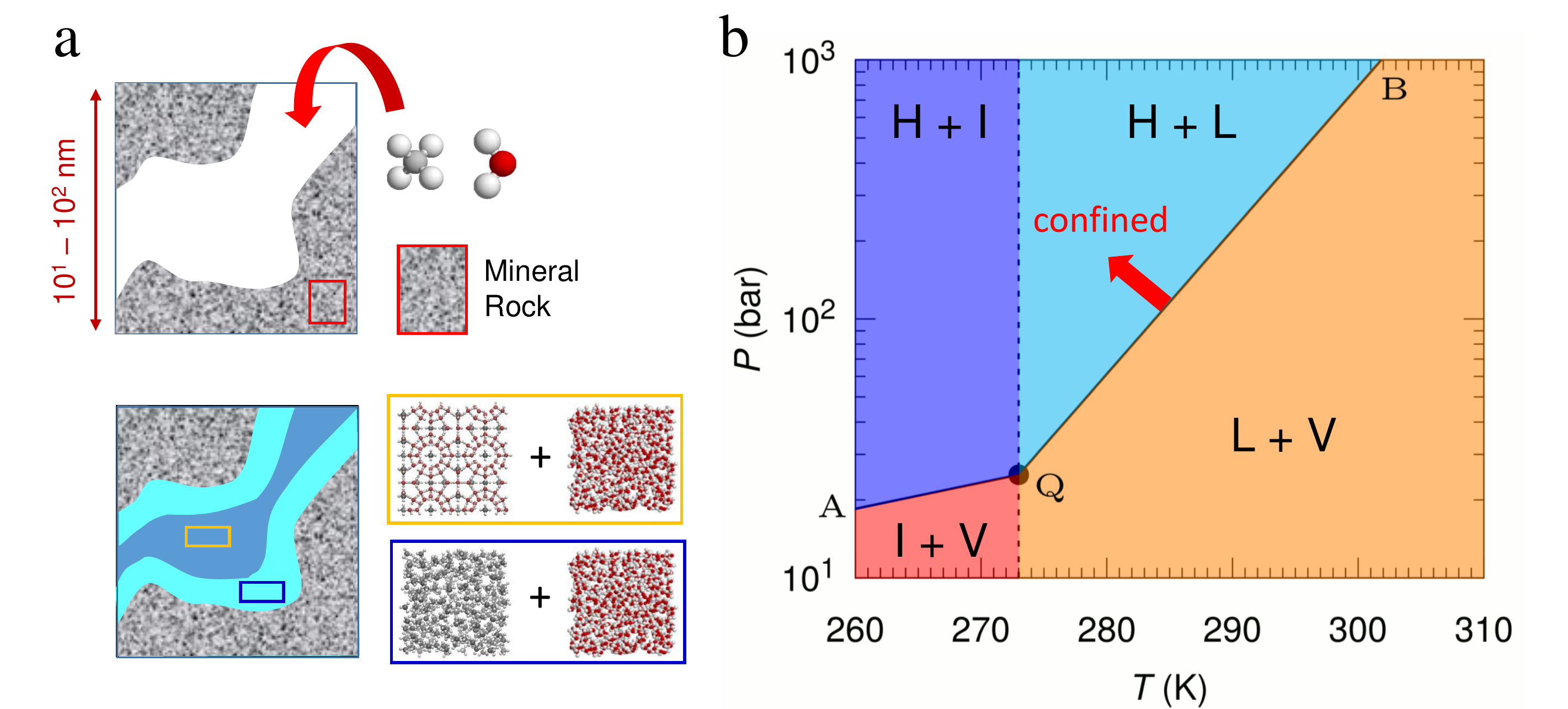}
        \caption{Phase equilibrium of methane hydrate--liquid water in nanoporous rocks. (a) Illustration of confined liquid water (cyan) and methane hydrate (blue) with molecular configurations of methane hydrate, liquid water, and methane. The red and white spheres are the oxygen and hydrogen atoms of water. The gray spheres are the methane molecules which are located inside the  cages formed by water molecules in methane hydrate. (b) Bulk pressure--temperature ($P$--$T$) phase diagram for ice (I), liquid (L), hydrate (H), and vapor (V). The dashed line denotes the ice melting line while the solid line A-Q-B corresponds to the hydrate melting line: for a given pressure $P$, H exists at low $T$ while V exists at high $T$. Confinement is expected to shift the melting point $P$, $T_f(P)$ to lower $T$ and larger $P$.}
        \label{fig:Figure01}
    \end{SCfigure*}

The phase diagram of bulk methane hydrate has been the subject of intense research to identify $P/T$ coexistence conditions as well as to determine formation/dissociation mechanisms~\cite{hall_evidence_2016, duan_theinfluence_2011}. In contrast, the case of methane hydrate confined in porous media remains unclear by many aspects with important questions left regarding the role of confinement and surface forces~\cite{casco_methane_2015, wang_direction_2015, ghosh_clathrate_2019}. Yet, in nature, abundant methane hydrate resources are found in the porosity of rocks and minerals such as in marine sediments, silica sands, permafrost, etc. so that understanding confinement or surface-induced shifts in the phase stability, formation/dissociation kinetics, and composition of gas hydrates is of utmost importance (Fig.~\ref{fig:Figure01}). Many experimental observations have shown that confinement in pores shifts the liquid--hydrate--vapor (L--H--V) phase boundaries to higher $P$ and/or lower $T$ as a result of capillary forces~\cite{borchardt_illuminating_2016, borchardt_methane_2018, casco_experimental_2019, seo_methane_2002, uchida_effects_2002}. For both organic (e.g. carbonaceous) and mineral (e.g. siliceous) porous environments, such experiments have shown that the shift in melting point qualitatively follows the Gibbs--Thomson equation with $\Delta T_f$ scaling linearly with the reciprocal of the pore size $D_p$, i.e. $\Delta T_f \sim 1/D_p$ [e.g. Refs.~\cite{seshadri_measurements_2001, anderson_experimental_2003} and Ref.~\cite{borchardt_methane_2018} for a recent review]. This classical macroscopic behavior is found to be reminiscent even at the nanometer scale \cite{deroche_reminiscent_2019} where the validity of macroscopic concepts such as surface tension, enthalpy of melting, etc. remains questionable [Fig.~\ref{fig:Figure01}(b)]. With this respect, while significant efforts are devoted to verifying the scaling law $\Delta T_f \sim 1/D_p$ using molecular simulation for nanoconfined methane hydrate~\cite{chakraborty_amonte_2012}, the question of its quantitative validity relying on robust calculations  of the pore/liquid and pore/hydrate surface tensions and other thermodynamic parameters has not been considered. In fact, while the question of a contact angle $\theta$ between solid and liquid phases at a pore surface remains to be addressed, many experimental and theoretical works assume that there is a wetting parameter $\omega = \gamma_\textrm{LH} \cos\theta$ that controls the shift in the melting point $\Delta T_f \sim \omega/D_p$ (with the usual assumption $\cos\theta \sim -1$). Beyond confinement-induced shifts in hydrate stability, the formation and dissociation kinetics of these complex compounds remain to be investigated in detail. There are abundant literature data indicating that confinement leads to faster kinetics compared to their bulk counterpart~\cite{casco_methane_2015, borchardt_methane_2018, linga_enhanced_2012} but the underlying microscopic mechanisms remain to be unraveled. From a theoretical viewpoint, much work has been done to elucidate the molecular routes that lead to the nucleation of bulk hydrates~\cite{walsh_microsecond_2009, yuhara_nucleation_2015, knott_homogeneous_2012, arjun_unbiased_2019, liang_nucleation_2019} but the case of confined hydrates has not been considered. Several authors have used molecular modeling to investigate the growth of an already formed methane or carbon dioxide hydrate in the vicinity of a surface~\cite{kang_kinetics_2009, bai_microsecond_2011, yan_molecular_2016, liang_crystal_2011, bagherzadeh_influence_2012} but microscopic nucleation remains largely unexplored.

Here, we present a theoretical study of confined methane hydrate to elucidate the molecular mechanisms responsible for their reduced phase stability and faster formation/dissociation kinetics. By relying on statistical mechanics, such molecular simulations -- using either the direct coexistence method (DCM) or free energy calculations -- capture phase coexistence between confined liquid water and methane hydrate. In particular, our approach does not assume any thermodynamic modeling framework or kinetic pathway. We first extend the direct coexistence method to the Grand Canonical ensemble (open ensemble) to  account for three phase coexistence (L--H--V) involved in gas hydrate formation/dissociation. Using such atom-scale simulations, we show that the melting point depression for confined hydrate is quantitatively consistent with the Gibbs--Thomson equation if the pore/liquid and pore/hydrate surface tensions are used (here, values were calculated using independent molecular simulations). This macroscopic expression is shown to be a simplified thermodynamic approach that neglects the effect of pressure/temperature and methane chemical potential. In agreement with experimental data, the formation and dissociation of confined methane hydrate as described in our mesoscopic description are faster than for bulk methane hydrate. By considering different nucleation paths (bulk homogeneous nucleation, external surface nucleation, and in-pore confined nucleation), we identify the critical nucleus leading to the formation of methane hydrate at the surface of the host porous material. This mesoscale thermodynamic approach unravels a crossover in the nucleation process; depending on temperature, the formation of methane hydrate within the porosity involves a critical nucleus that corresponds either to a hemispherical cap or a bridge nucleus. In both cases, within the classical nucleation theory, we derive a single expression for the nucleation rate $1/\tau$ which is found to scale with the reciprocal of the material specific surface area or, equivalently, with the pore size $D_p$, $1/\tau \sim D_p$.

\section*{Results}

\begin{SCfigure*}[\sidecaptionrelwidth][t]
        \centering
        \includegraphics[width=12cm]{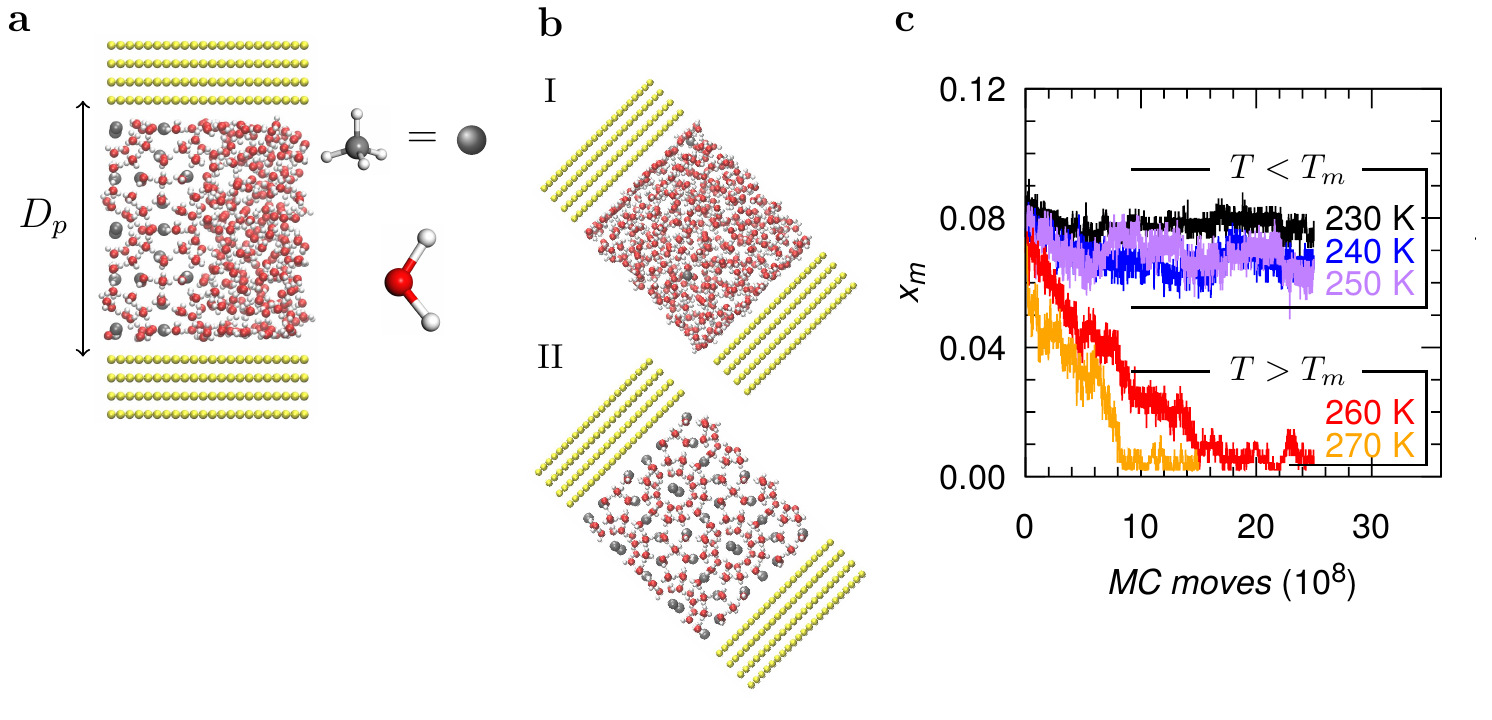}
        \caption{Probing methane hydrate stability in nanopores. (a) Molecular modeling of phase coexistence in a pore of width $D_p \sim 2.86$ nm. Hydrate and liquid water are located on the left and right, respectively. The red and white spheres are the water oxygen and hydrogen atoms. The gray spheres are methane molecules inside the hydrogen-bonded cages formed by water. The yellow spheres are the solid atoms distributed according to a face-centered square structure. The box dimensions are $L_x = L_y \sim 2.38$ nm. (b) Starting from coexistence in (a), the system evolves towards liquid water or methane hydrate depending on $T$: (I) for $T>T_f$, hydrate melts, (II) for $T<T_f$, liquid water crystallizes into hydrate ($T_f$ is the melting point of confined  hydrate). (c) Change in methane mole fraction $x_{m}$ during the GCMC simulations at different $T$.}
        \label{fig:Figure02}
    \end{SCfigure*}

\textbf{Melting point in confinement.} Fig.~\ref{fig:Figure02}(a) illustrates the direct coexistence method (DCM) which is extended here to the Grand Canonical ensemble to investigate the effect of confinement on L--H--V equilibrium (SI \textit{Text}). A typical molecular configuration of methane hydrate coexisting with liquid water in a slit pore of a size $D_p$ is prepared ($D_p \sim 1.67$ nm, 2.86 nm, 5.23 nm, and 7.61 nm are considered). Using this starting configuration, for a given pressure $P$, Monte Carlo simulations in the Grand Canonical ensemble (GCMC) are performed for different temperatures $T$ to determine the melting temperature $T_f$ as follows. Methane hydrate melts into liquid water for $T>T_f$ [I in Fig.~\ref{fig:Figure02}(b)] while liquid water crystallizes into methane hydrate for $T < T_f$ [II in Fig.~\ref{fig:Figure02}(b)]. The Grand Canonical ensemble ensures that L--H--V coexistence is simulated \textit{de facto}~\cite{coasne_freezing_2007, coasne_effect_2009}; because the system is in equilibrium with an infinite reservoir at chemical potentials corresponding to bulk liquid water $\mu_w(P,T)$ and methane vapor $\mu_m(P,T)$ at the same $P$ and $T$, DCM in this ensemble is equivalent to simulating a three phase coexisting system [$\mu_m(P,T)$ and $\mu_w(P,T)$ are taken from~\cite{jin_molecular_2017}]. 

\begin{SCfigure*}[\sidecaptionrelwidth][t]
        \centering
        \includegraphics[width=10cm]{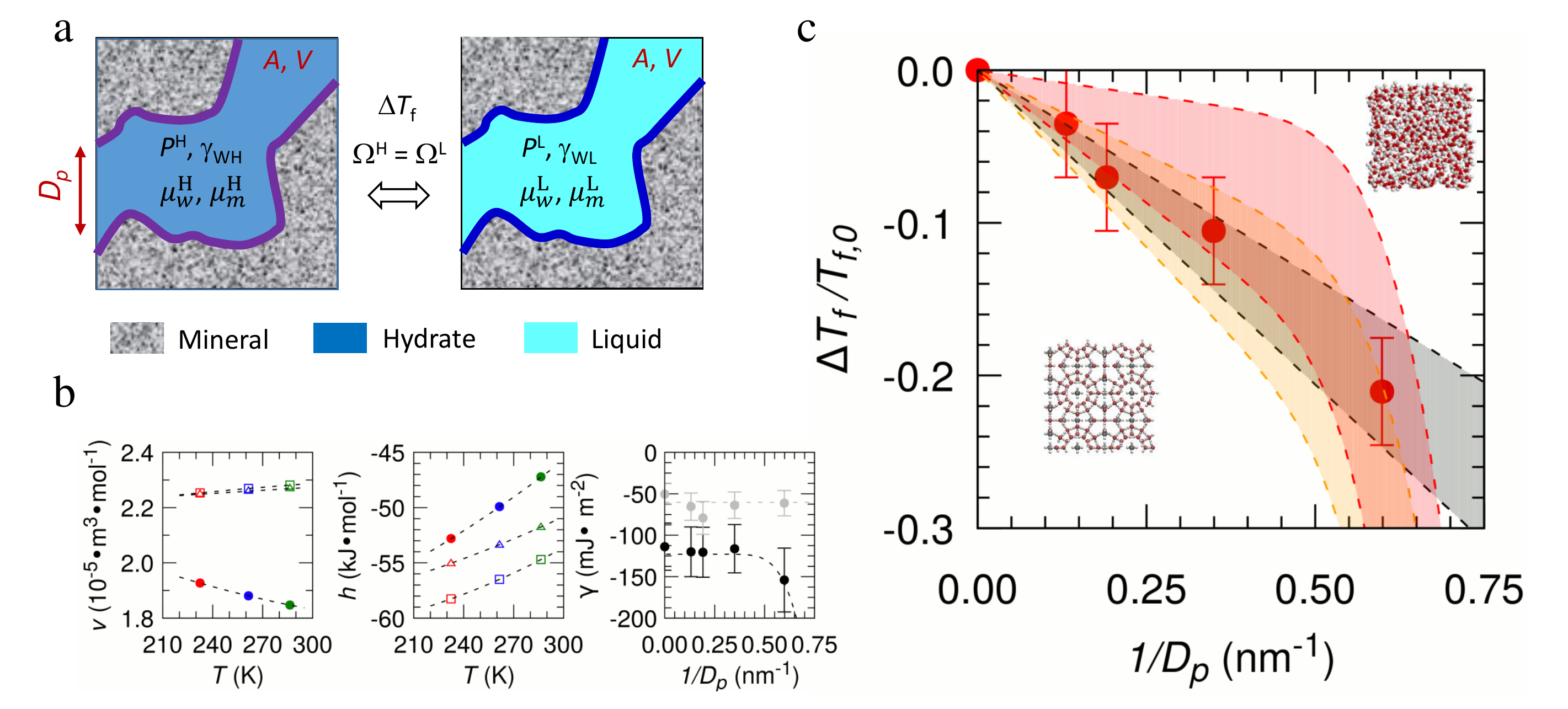}
        \caption{Extended Gibbs--Thomson effect. (a) L--H--V equilibrium in pores: liquid water (cyan) and hydrate (blue) are confined in a porous material (gray) of pore width $D_p$, surface area $A$, and volume $V$. $\Omega^\textrm{H}$ and $\Omega^\textrm{L}$ are the hydrate and water grand potentials which are at pressures $P^\textrm{H}$ and $P^\textrm{L}$. (b) Molar volume $v$ and enthalpy $h$ for liquid water (circles), full-occupancy hydrate (squares), and empty-occupancy hydrate (triangles). These parameters are determined along bulk L--H--V equilibrium at 233 K/1 atm (red), 262 K/10 atm (blue), and 286 K/100 atm (green). The right panel shows the pore/liquid $\gamma_{\textrm{WL}}$ (black) and pore/hydrate $\gamma_{\textrm{WH}}$ (gray) surface tensions as a function of $1/D_p$ at the corresponding melting point. (c) Melting shift $\Delta T_f/T_{f,0}$ for confined hydrate $T_f$ with respect to bulk $T_{f,0}$ at $P=100$ atm as a function of $1/D_p$. The solid circles are obtained from direct coexistence method, while the dashed lines are determined using the Gibbs--Thomson equation with different approximations level (see text). For each model, a range of values -- denoted by the color shaded area -- is indicated to account for error bar.}
        \label{fig:Figure03}
    \end{SCfigure*}

Fig.~\ref{fig:Figure02}(c) shows the methane mole fraction $x_m$ for the slit pore with $D_p \sim 2.86$ nm at different $T$ in the course of GCMC simulations (data for other $D_p$ are given in Figs. S1 and S2); $x_m$ decreases to 0 as the system melts for $T \geq 260$ K while $x_m$ increases upon hydrate formation for $T \leq 250 $ K. These data show that $T_f=255\pm5$ K for $D_p \sim 2.86$ nm, which is lower than the bulk melting point taken at the same $P=100$ atm ($T_{f,0}=285\pm5$ K). This result, which is in qualitative agreement with experimental data on confined methane hydrate, suggests that the pore/hydrate surface tension $\gamma_\textrm{WH}$ is larger than the pore/liquid surface tension $\gamma_\textrm{WL}$. The red circles in Fig.~\ref{fig:Figure03}(c) show the melting point depression $\Delta T_f/T_{f,0}$ obtained using  DCM as a function of $1/D_p$ (as shown below, these data are consistent with free energy calculations). In qualitative agreement with the Gibbs--Thomson equation, the scaling $\Delta T_f \sim 1/D_p$ is observed even for such small nanopores. While many experimental and simulation works have validated this scaling, the quantitative verification of the Gibbs--Thomson equation applied to such ultra-confinement has been hampered by limitations in assessing pore/hydrate and pore/liquid surface tensions. The effect of surface wettability is assessed here by changing the LJ energy parameter $\epsilon^\prime$ of the pore/hydrate and pore/liquid pair interactions. $\epsilon^\prime = \epsilon/2$, $\epsilon/3$, $\epsilon/4$, $2 \epsilon$, $3 \epsilon$, and $4 \epsilon$ (where $\epsilon$ is the original LJ energy parameter in Table S1) are adopted to mimic stronger or weaker surface interactions. As shown in Fig. S3, $T_f$ remains constant as  $\epsilon^\prime$ is varied. This result can be explained using the Gibbs--Thomson equation [\eqref{eq:gibbsthomsonequationA}] which will be discussed in detail below. The shift $\Delta T_{f}/T_{f,0}$ is proportional to the surface tension difference $\Delta \gamma = \gamma_\textrm{WL} - \gamma_\textrm{WH}$. At constant $T$ and $P$, a Taylor expansion for $\Delta \gamma$ leads to $\Delta \gamma(\epsilon^\prime) \sim \Delta\gamma(\epsilon) + \partial\Delta\gamma(\epsilon)/\partial\epsilon \times (\epsilon^\prime-\epsilon)$. Surface interactions are found to amount for $<5\%$ of the energy with negligible impact on $\Delta \gamma$. As a result, $\Delta \gamma(\epsilon^\prime) \sim \Delta \gamma(\epsilon)$ so that wettability considered here leads to the same melting shift.

\noindent\textbf{Gibbs--Thomson formalism.} Independent molecular simulations were used to determine parameters to assess the quantitative validity of the Gibbs--Thomson equation. Capillary crystallization is described at the macroscopic level using this equation -- which is analogous to the Kelvin equation but for liquid/solid transitions. Such thermodynamic model relies on Laplace equation (which links the pressure difference in the confined crystal and liquid to their surface tension) combined with the chemical potential equality between the different phases. Two important assumptions are usually made when deriving the Gibbs--Thomson equation: (1) the crystal and liquid have the same molar volume ($v_\textrm{C} \sim v_\textrm{L}$) and (2) Young equation holds for liquid/solid systems with a contact angle $\theta$ ($\gamma_\textrm{WL} - \gamma_\textrm{WC} = \gamma_\textrm{LC}\cos\theta$). While the robustness of the first assumption can be assessed, the second assumption is key as the concept of solid/liquid contact angle in confinement remains unclear. A third assumption, which only pertains to multi-component phases such as hydrates, consists of neglecting the impact of the lowest mole fraction component on stability. For methane hydrate, this assumption consists of deriving phase stability without considering the chemical potential contribution from methane vapor. In what follows, we formally derive the Gibbs--Thomson equation without invoking the contact angle and extends its applicability to binary solids by including the methane contribution into the stability condition. Then, using independent molecular simulations, we estimate the different ingredients to discuss the validity of the Gibbs--Thomson equation  when applied to confined hydrate (all derivation steps can be found in SI \textit{Text}).

Let us consider two phases $\Phi$, methane hydrate ($\Phi=$ H) and liquid water with solubilized methane ($\Phi=$ L), in a slit pore of size $D_{p}$, surface area $A$, and pore volume $V = D_pA$ [Fig.~\ref{fig:Figure03}(a)]. These two phases are in equilibrium with an infinite bulk reservoir which imposes the water and methane chemical potentials $\mu_w$ and $\mu_m$, and temperature $T$. For hydrate/liquid equilibrium in confinement, considering the grand potential $\Omega^\Phi = -P^\Phi V + 2\gamma_{\textrm{W}\Phi}A$ for each confined phase ($\Phi =$ L, H), grand potential equality leads to the Laplace equation:  
\begin{equation} \label{eq:laplaceequation}
    P^\textrm{L} - P^\textrm{H} = 2\left(\gamma_\textrm{WL} - \gamma_\textrm{WH}\right)/D_{p}
\end{equation}
where the factor 2 accounts for the two surfaces in the slit geometry. $\gamma_{\textrm{WH}}$ and $\gamma_{\textrm{WL}}$ are the pore/hydrate and pore/liquid surface tensions at the confined melting point ($\mu_w, \mu_m, T_f$). In confinement, as shown in SI \textit{Text}, the pressures $P^\Phi$ at the melting point $(\mu_w,\mu_m,T_f)$ can be expressed using a Taylor expansion around the bulk melting point ($\mu_{w,0}$, $\mu_{m,0}$, $T_{f,0}$):
\begin{equation} \label{eq:pressureconfinement}
    \begin{split}
    & P^\textrm{H} = P_0 + \frac{1}{v_0^\textrm{H}}\left( \Delta T_f s_0^\textrm{H} + \Delta \mu_w + \Delta \mu_m \frac{x_{m,0}^\textrm{H}}{1-x_{m,0}^\textrm{H}} \right) \\
    & P^\textrm{L} = P_0 + \frac{1}{v_0^\textrm{L}}\left( \Delta T_f s_0^\textrm{L} + \Delta \mu_w + \Delta \mu_m \frac{x_{m,0}^\textrm{L}}{1-x_{m,0}^\textrm{L}} \right)
    \end{split}
\end{equation}
where $P^{\Phi} = P^{\Phi}(\mu_w,\mu_m,T_f)$ is the pressure of phase $\Phi$ at ($\mu_{w}$, $\mu_{m}$, $T_f$) and $P_0 = P_0^\textrm{H}(\mu_{w,0},\mu_{m,0},T_{f,0}) = P_0^\textrm{L}(\mu_{w,0},\mu_{m,0},T_{f,0})$ is the pressure at  ($\mu_{w,0}$, $\mu_{m,0}$, $T_{f,0}$) [corresponding to $D_p \sim \infty$ in the Laplace equation]. $\Delta T_f = T_f - T_{f,0}$ is the melting temperature shift of confined methane hydrate with respect to its bulk counterpart. $\Delta \mu_i = \Delta \mu_{i}^{\Phi}=\mu_i^{\Phi}-\mu_{i,0}^{\Phi}$ is the difference of chemical potential $\mu_{i}^{\Phi}$ of species $i$ [$i=$ methane ($m$), water ($w$)] for phase $\Phi$ at $T_f$ (at this point $\mu_{i}^\textrm{H} = \mu_{i}^\textrm{L} = \mu_i$ because of phase coexistence in confinement while also in equilibrium with the external phase) and $T_{f,0}$ (at this point $\mu_{i,0}^\textrm{H} = \mu_{i,0}^\textrm{L} = \mu_{i,0}$ by definition of bulk phase equilibrium). $s_0^{\Phi}/v_0^{\Phi} = \partial P/\partial T \left(\mu_{w,0},\mu_{m,0},T_{f,0}\right)$ is the molar entropy $s_0^{\Phi}$ (note that $s$ is the total entropy which includes both methane and water contributions) divided by the molar volume $v_0^{\Phi}$ for phase $\Phi$ at ($\mu_{w,0}$, $\mu_{m,0}$, $T_{f,0}$). $1/v_0^{\Phi} = \partial P/\partial \mu_w \left(\mu_{m,0},\mu_{w,0},T_{f,0}\right)$ is the reciprocal of the molar volume for phase $\Phi$ at ($\mu_{w,0}$, $\mu_{m,0}$, $T_{f,0}$). As established in Eq. (S4), deriving $\Omega$ shows that  $N_{m,0}^\Phi/N_{w,0}^\Phi v_0^\Phi=\partial P/\partial\mu_m \left(\mu_{m,0},\mu_{w,0},T_{f,0}\right)$ is the ratio of the number of methane and water molecules divided by the molar volume $v_0^\Phi$ for phases $\Phi$ at ($\mu_{w,0}$, $\mu_{m,0}$, $T_{f,0}$). By noting that $N_{m,0}^\Phi/N_{w,0}^\Phi = x_{m,0}^{\Phi}/(1-x_{m,0}^{\Phi})$ where $x_{m,0}^{\Phi}$ is the methane mole fraction in phase $\Phi$ at ($\mu_{w,0}$, $\mu_{m,0}$, $T_{f,0}$) and that $x_{m,0}^\textrm{L} \sim 0$ for the liquid phase, Eqs.~(\ref{eq:laplaceequation}) and~(\ref{eq:pressureconfinement}) lead to:
\begin{equation}\label{eq:gibbsthomsonequationA}
    \begin{split}
    \frac{\Delta T_f}{T_{f,0}} & = \frac{v_0^{\textrm{H}}}{\Delta h_{f,0}} \left[ \left(\gamma_\textrm{WL} - \gamma_\textrm{WH}\right)\frac{2}{D_{p}} \right.\\
    & \left. + \left(\frac{v_0^{\textrm{L}}}{v_0^{\textrm{H}}}-1\right)\left(P^\textrm{L}-P_{0}\right) + \frac{\Delta \mu_m}{v_0^{\textrm{H}}} \frac{x_{m,0}^{\textrm{H}}}{1-x_{m,0}^{\textrm{H}}} \right]
    \end{split} 
\end{equation}
where $\Delta h_{f,0}=\left(s_{f,0}^\textrm{L}-s_{f,0}^\textrm{H}\right)T_{f,0}$ is the bulk molar enthalpy of melting at ($T_{f,0}$, $P_0$). At this stage, it is often assumed that (1) the hydrate and liquid molar volumes are similar ($v_0^{\textrm{H}} \sim v_0^{\textrm{L}}$) and that (2) the methane chemical potential contribution is negligible ($\Delta \mu_m \sim 0$). With these approximations, one arrives at the Gibbs--Thomson equation which relates $\Delta T_f$ to $D_p$:
\begin{equation}\label{eq:gibbsthomsonequationB}
    \frac{\Delta T_f}{T_{f,0}} = \frac{2v_0^{\textrm{H}} \left( \gamma_\textrm{WL} - \gamma_\textrm{WH} \right)}{\Delta h_{f,0}} \frac{1}{D_{p}}
\end{equation}
Introducing Young equation, $\gamma_\textrm{WL} - \gamma_\textrm{WH} = \gamma_\textrm{LH} \cos\theta$ with $\theta=\pi$, \eqref{eq:gibbsthomsonequationB} leads to the classical formulation: $\Delta T_f/T_{f,0} = -2\gamma_\textrm{LH}v_0^{\textrm{H}}/ \Delta h_{f,0}D_{p}$. In contrast to this simplified equation, \eqref{eq:gibbsthomsonequationA} is a revised version that accounts for the effect of methane gas in the free energy balance and the significant density difference between the liquid and hydrate.

In what follows, we determine the following parameters using independent molecular simulations to check the quantitative validity of \eqref{eq:gibbsthomsonequationA}: molar volumes $v^{\textrm{H}}$ and $v^{\textrm{L}}$, molar enthalpy of melting  $\Delta h_f = \left(h_{w}^{\textrm{L}} + h_{m}^{\textrm{V}}\right) - h_{m,w}^{\textrm{H}}$ from the methane hydrate, liquid water, and methane vapor molar enthalpies. Fig.~\ref{fig:Figure03}(b) shows $v^\textrm{H}$, $v^\textrm{L}$, $h^\textrm{H}$, and $h^\textrm{L}$ at different bulk equilibrium conditions: (233 K, 1 atm), (262 K, 10 atm), and (286 K, 100 atm). We obtain $\Delta h_f=8.35$ kJ$\cdot$mol\textsuperscript{-1}, $v^\textrm{L} = 1.85\times10^{-5}$ m\textsuperscript{3}$\cdot$mol\textsuperscript{-1}, and $v^\textrm{H}=2.28\times10^{-5}$ m\textsuperscript{3}$\cdot$mol\textsuperscript{-1} at $T=286$ K and $P=100$ atm. Such enthalpy of melting $\Delta h_f$ leads to an entropy of melting $\Delta s_f=\Delta h_f/T_{f,0}=29.3$ J$\cdot$K\textsuperscript{-1}$\cdot$mol\textsuperscript{-1} which is comparable to that reported in~\cite{jacobson_can_2011}. In addition to these parameters, the extended Gibbs--Thomson equation requires to determine $\gamma_\textrm{WH}$ and $\gamma_\textrm{WL}$. Here, we use the Irving-Kirkwood approach to determine $\gamma_\textrm{WH}$ and $\gamma_\textrm{WL}$ as described in SI \textit{Text}. Fig.~\ref{fig:Figure03}(b) shows $\gamma_\textrm{WH}$ and $\gamma_\textrm{WL}$ as a function of $1/D_p$. Each calculation was not carried out at constant $T$/$P$ but at the temperature corresponding to the melting point upon confinement. This allows quantitative assessment of \eqref{eq:gibbsthomsonequationA} since all surface tensions should be taken at the corresponding melting point ($\gamma_\textrm{WH}^0$ and $\gamma_\textrm{WL}^0$ were also estimated at the bulk melting point as often invoked when applying the Gibbs--Thomson equation). 

Fig.~\ref{fig:Figure03}(c) compares the predictions from the different versions of the Gibbs--Thomson equation with the DCM results. The classical version of the Gibbs--Thomson equation -- which neglects the liquid/solid density difference and methane chemical potential -- is found to reasonably describe the shift in melting point observed using molecular simulation (see comparison between the red circles and the gray shaded area). This result is consistent with previous experimental and molecular simulation results showing a qualitative scaling $\Delta T_f/T_{f,0} \sim  1/D_p$~\cite{chakraborty_amonte_2012, anderson_experimental_2003, uchida_effects_2002, seshadri_measurements_2001, seo_methane_2002}. As discussed above, possible refinement to the classical Gibbs--Thomson equation consists of using the different surface tensions $\gamma$ at the confined melting point $(T_f, P_0)$ instead of their values $\gamma_0$ at the bulk coexistence point ($T_{f,0}, P_0)$. As shown in Fig.~\ref{fig:Figure03}(c), upon using the correct surface tensions, the classical Gibbs--Thomson equation also predicts quantitatively the melting point in confinement as a function of pore size (comparison between red circles and the orange shaded area). However, considering the error bar on the different data sets (i.e. molecular  simulation results and predictions based on the estimated thermodynamic parameters), the use of refined thermodynamic surface tensions is found to be of little improvement compared to the classical Gibbs--Thomson equation with values taken at the bulk melting point. This further supports previous works in the field where the Gibbs--Thomson equation is usually used with a single set of parameters for all  thermodynamic variables (which leads to a linear scaling $\Delta T_f \sim 1/D_p$ while the use of variable surface tensions provides a non-linear scaling). 

As for the extended Gibbs--Thomson equation, a few remarks are in order. First, by construction, because all simulations were carried out along the bulk liquid-hydrate phase coexistence, the confined liquid and hydrate are in equilibrium at all temperatures with a water/methane reservoir at the bulk coexistence pressure (i.e. $P^{\textrm{L}} = P^{\textrm{H}} = P^0$). This implies that the second term in \eqref{eq:gibbsthomsonequationA} is equal to zero. As a result, by comparing the rest of this extended equation with the classical Gibbs--Thomson equation in \eqref{eq:gibbsthomsonequationB} shows only one important difference: the methane chemical potential contribution. Fig.~\ref{fig:Figure03}(c) shows that the Gibbs--Thomson equation using the surface tensions at the confined melting point in combination with the chemical potential contribution as described in \eqref{eq:gibbsthomsonequationA} correctly predicts the melting point depression observed using molecular simulation (comparison between red circles and the red shaded area). In fact, the contribution arising from the corrected surface tensions and methane chemical potential more or less cancel out so that the resulting prediction is close to the classical equation with uncorrected surface tensions.

    \begin{SCfigure*}[\sidecaptionrelwidth][ht]
        \centering
        \includegraphics[width=10.cm]{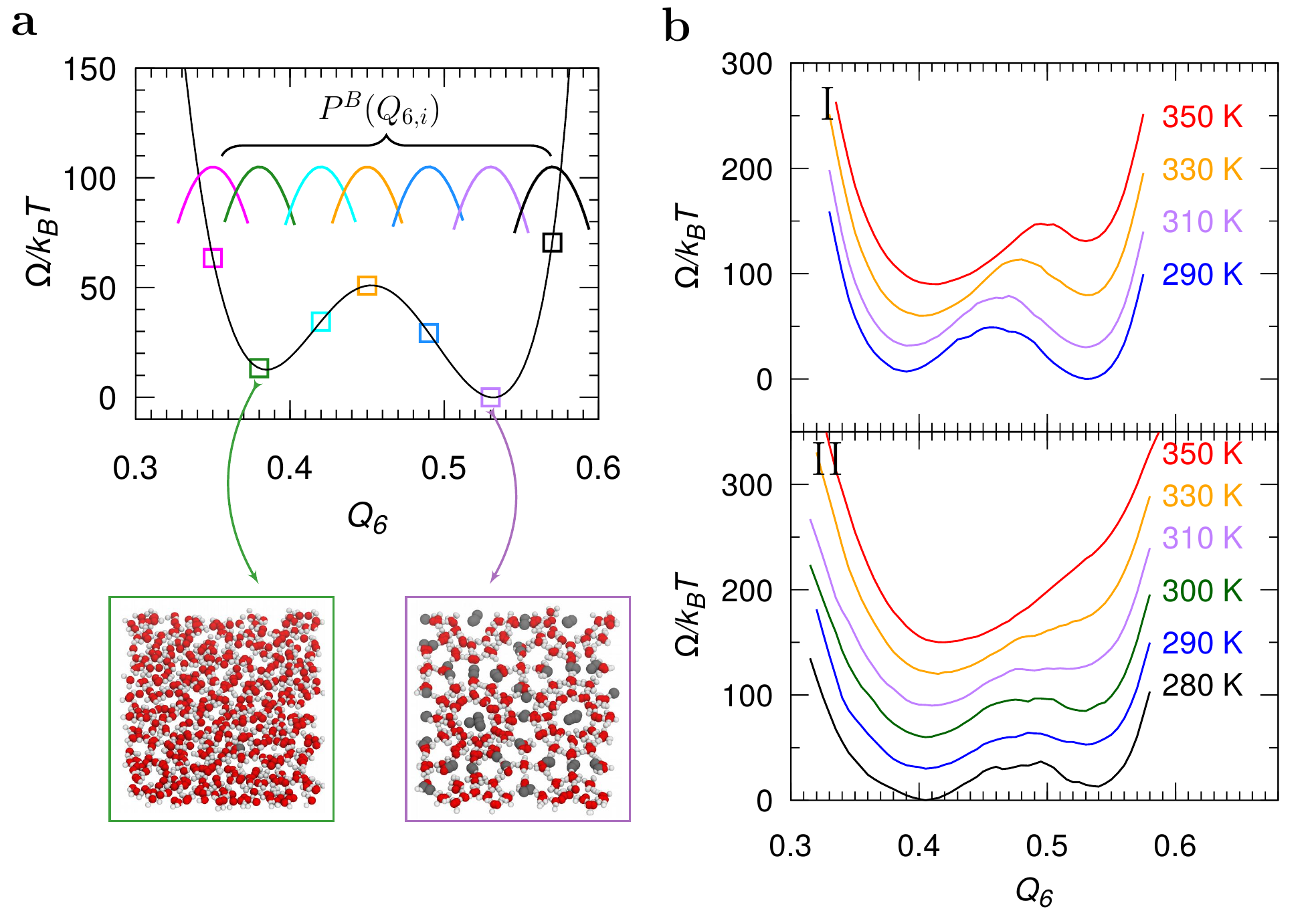}
        \caption{Free energy calculations for methane hydrate and liquid water. (a) Schematic illustration of the umbrella sampling technique: the black solid line in the graph shows the free energy $\Omega$ as a function of the local order parameter $Q_6$. In the umbrella sampling technique, a biased potential energy $U^B(Q_6)$ (see \textit{text}) is used to force the system to explore configurations around a given $Q_6$ (as indicated by the color squares which correspond to different $Q_6$). Methane hydrate ($Q_6\sim0.53$) and liquid water ($Q_6\sim0.38$) correspond to the two minima observed in $\Omega(Q_6)$. The biased simulations provide estimates for the biased probability distribution $P^B(Q_{6})$. The unbiased free energy $\Omega(Q_6)$ can be computed using the biased distribution minus the biased potential energy: $\Omega(Q_6)=-k_BT\ln [P^B(Q_6)]-U^B(Q_6)$. The red and white spheres are the oxygen and hydrogen atoms of water, respectively. The gray spheres are the methane molecules. (b) Free energy profile $\Omega(Q_6)$ calculated using the umbrella sampling technique for bulk (I) and confined (II) transitions at different $T$ (as labelled in the graph). All free energies are normalized to the thermal energy $k_BT$.} 
        \label{fig:Figure04}
    \end{SCfigure*}

\noindent\textbf{Free energy and phase stability.} To probe the formation and dissociation kinetics of confined methane hydrate, we combined GCMC simulations with free energy calculations (details are provided in the Methods section). Using the umbrella-sampling technique, one can determine the free energy profile $\Omega(x)$ as a function of an order parameter $x$, which describes the transition from the liquid to the crystal phases [Fig.~\ref{fig:Figure04}(a)]. The local order parameter $x=Q_6$ was used to identify methane hydrate ($Q_6\sim0.55$) and liquid water ($Q_6\sim0.35$). A perfect methane hydrate of dimensions $L_x=L_y=L_z\sim 2.36$ nm was prepared for bulk and confined methane hydrate in a slit pore of width $D_p \sim 2.86$ nm. Then, a harmonic potential $U^B(Q_6)$ is added to the energy calculated in the GCMC simulations to force the system to sample states corresponding to the given $Q_{6}$ value [This strategy is illustrated in Fig.~\ref{fig:Figure04}(a) using the color squares]. Such a biased potential allows one to determine the probability distribution $P^B(Q_6)$ from the biased molecular simulations. Finally, the free energy profile is obtained by subtracting the biased potential energy from the biased probability distribution: $\Omega(Q_6) = -k_BT\ln [P^B(Q_6)] - U^B(Q_6)$.

Fig.~\ref{fig:Figure04}(b) shows the free energy $\Omega/k_BT$ for the bulk (I) and confined (II, $D_p \sim 2.86$ nm) hydrate/liquid  transitions as a function of the local order parameter $Q_6$ at different $T$. For the bulk transition [I in Fig.~\ref{fig:Figure04}(b)], methane hydrate is found to be stable for $T<T_{f,0}$. For the confined transition [II in Fig.~\ref{fig:Figure04}(b)], all free energy calculations are performed above the expected melting temperature $T_f$ of confined methane hydrate as the umbrella sampling technique at lower $T$ failed to converge (as discussed below, this is due to the slow formation/dissociation kinetics for methane hydrate). As expected, liquid water is the favorable phase at these $T$; indeed, free energy difference between methane hydrate and liquid water at these $T$ is negative. To estimate the melting temperature of bulk and confined methane hydrate, the free difference $\Delta\Omega$ as a function of temperature $T$ is shown in Fig. S7; We find that $\Delta\Omega$ depends linearly on $T$ with the melting temperature corresponding to the temperature for which $\Delta\Omega \sim 0$. This extrapolation leads to $T_{f,0}=302$ K and $T_f=257$ K for $D_p \sim 2.86$ nm which are in good agreement with the DCM.

\begin{SCfigure*}[\sidecaptionrelwidth][ht]
        \centering
        \includegraphics[width=10cm]{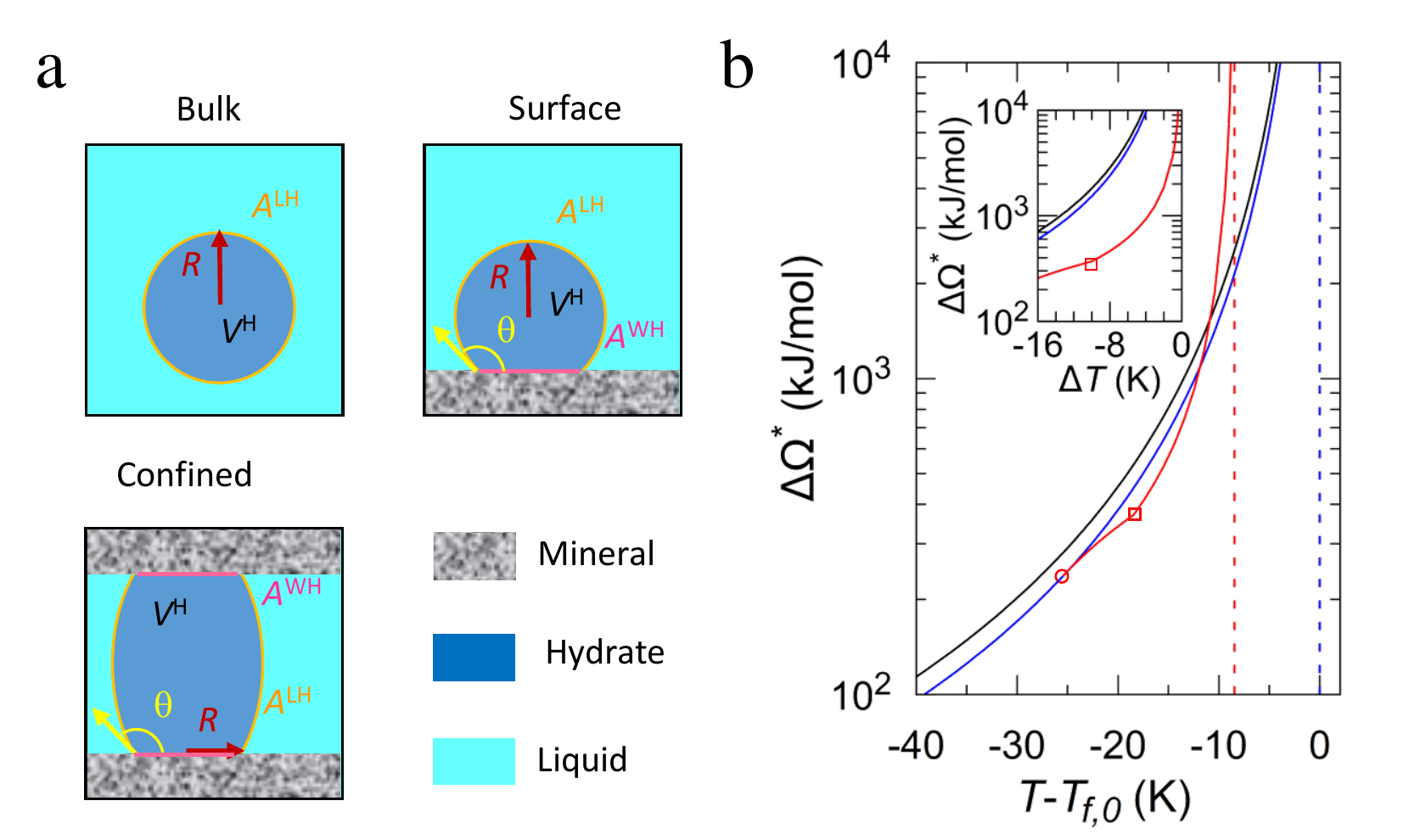}
        \caption{Nucleation mechanisms for confined methane hydrate. (a) Schematic nucleus along possible mechanisms: bulk, surface, and confined nucleation. Methane hydrate and liquid are shown as dark and light blue phases while the solid surface is shown in grey. $V^\textrm{H}$ and $R$ are the volume and radius of the nucleus while $A^\textrm{LH}$ (orange line) and $A^\textrm{WH}$ (pink line) are the hydrate/liquid and pore/hydrate interfaces. $\theta$ is the contact angle where the liquid, hydrate, and solid intersect. (b) Nucleation free energy barriers $\Delta \Omega^\ast$ calculated using the mesoscopic model for bulk (black), surface (blue) and confined (red) nucleation as a function of the temperature shift $T - T_{f,0}$ with respect to $T_{f,0}$. The blue and red dashed lines indicate the formation temperature for bulk $T_f$ and confined $T_{f,0}$ methane hydrate, respectively. The inset shows the same data $\Delta\Omega^\ast$ but as a function of the temperature shift with respect to the formation temperature: $\Delta T = T - T_{f,0}$ for bulk/surface nucleation and $\Delta T = T - T_f(D_p)$ for confined nucleation. $D_p = 2.86$ nm is used for confined nucleation and all calculations are performed with $\gamma_\textrm{LH} = 32$ mJ/m$^2$ as estimated in Ref.~\cite{mirzaeifard_multiscale_2019}.}
        \label{fig:Figure05}
    \end{SCfigure*}

\noindent\textbf{Formation and dissociation kinetics.} Despite its robustness, umbrella sampling fails to provide a rigorous estimate for the nucleation barrier $\Delta \Omega^\ast$ because the inferred value depends on the chosen order parameter and finite system size. To apprehend the nucleation process involved in the formation and dissociation of confined methane hydrate, we extend our investigation using a \emph{mesoscale} description relying on thermodynamic ingredients identified in the molecular approach. As illustrated in Fig.~\ref{fig:Figure05}(a), considering a porous medium coexisting with a bulk phase made up of gas methane and liquid water, three possible mechanisms must be considered: (1) bulk nucleation in the external liquid phase, (2) surface nucleation at the external surface of the porous solid, and (3) confined nucleation within the porous solid. Bulk nucleation corresponds to homogeneous nucleation through the formation of a spherical nucleus which grows until it becomes unstable. Both surface and confined nucleation correspond to heterogeneous mechanisms that initiate at the solid surface. Surface nucleation involves the formation of a surface hemispherical cap that grows until it becomes unstable at the critical nucleus size. Confined nucleation corresponds to the formation and growth of a surface hemispherical cap which transforms into a bridge as its height reaches the opposite pore surface [Fig. S9(a)]. Using the classical nucleation theory~\cite{auer_quantitative_2004, aman_interfacial_2016}, the exact nucleus shape for each mechanism is obtained by minimizing the surface energy at fixed methane hydrate volume $V^\textrm{H}$~\cite{restagano_metastability_2000, lefevre_intrusion_2004}. Regardless of the nucleus geometry, its grand free energy can be expressed as $\Omega = -P^\textrm{H}V^\textrm{H} -P^\textrm{L}V^\textrm{L} + \gamma_\textrm{WH}A^\textrm{WH} + \gamma_\textrm{WL}A^\textrm{WL} + \gamma_\textrm{LH}A^\textrm{LH}$. The excess grand free energy $\Delta \Omega = \Omega - \Omega^\textrm{L}$ needed to generate such a nucleus with respect to the liquid phase ($\Omega^\textrm{L} = -P^\textrm{L}V + \gamma_\textrm{LW} A$) writes:
\begin{equation} \label{eq:DeltaOmega}
    \begin{split}
    \Delta\Omega  & =  (P^\textrm{L}-P^\textrm{H})V^\textrm{H} + (\gamma_\textrm{WH}-\gamma_\textrm{WL})A^\textrm{WH} + \gamma_\textrm{LH}A^\textrm{LH} \\
    & = \Delta P V^\textrm{H} - \gamma_\textrm{LH}\cos \theta A^\textrm{WH} + \gamma_\textrm{LH}A^\textrm{LH}
    \end{split}
\end{equation}
where $V = V^\textrm{H} + V^\textrm{L}$ and $A = A^\textrm{WH} + A^\textrm{WL}$ are the total pore volume and surface. The second equality is obtained by using $\Delta P = P^\textrm{L}-P^\textrm{H}$ and Young equation $\gamma_\textrm{WL} - \gamma_\textrm{WH} =  \gamma_\textrm{LH}\cos\theta$. For each formation mechanism, the nucleation barrier $\Delta \Omega^\ast$ corresponding to the critical nucleus is given by the maximum free energy along the nucleation path $\Delta \Omega(V^\textrm{H})$~\cite{altabet_effect_2017}.

Fig.~\ref{fig:Figure05}(b) compares the free energy barriers for bulk, surface, and confined nucleation at a given shift $T - T_{f,0}$ with respect to the bulk formation temperature $T_{f,0}$ (all analytical calculations can be found in SI \textit{text}). As expected, $\Delta \Omega^\ast$ diverges at $T_{f,0}$ for bulk and surface nucleation and at $T_{f}(D_p)$ for confined nucleation. The free energy barriers for the bulk and surface mechanisms are $\Delta \Omega^\ast = 4/3 \pi \gamma_\textrm{LH} {R^\ast}^2$ and $\Delta \Omega^\ast = \gamma_\textrm{LH} \pi {R^\ast}^2 (2 - 3 \cos \theta + \cos^3 \theta)/3$. As for confined nucleation, the exact critical nucleus and associated free energy barrier depend on $T - T_{f,0}$ (Fig. S9). For small $T - T_{f,0}$ [red line, right of the red square in Fig.~\ref{fig:Figure05}(b)], the critical nucleus corresponds to a hydrate bridge which forms as the hemispherical cap initiated at the pore surface becomes unstable. For intermediate $T - T_{f,0}$ [red line, between the red square and circle in Fig.~\ref{fig:Figure05}(b)], $\Delta \Omega^\ast$ corresponds to a critical nucleus consisting of the hemispherical cap becoming unstable as it reaches the opposite pore surface. For large $T - T_{f,0}$ [red line, left of red circle in Fig.~\ref{fig:Figure05}(b)], the critical nucleus consists of the hemispherical cap becoming unstable before reaching the opposite pore surface (as a result, in this temperature range, confined nucleation merges with surface nucleation). As shown in the inset Fig.~\ref{fig:Figure05}(b), if we consider $\Delta \Omega^\ast$ at a given $\Delta T$ with respect to $T_{f,0}$ or $T_f(D_p)$, $\Delta \Omega^\ast$ is much lower for confined nucleation than for bulk and surface nucleation. This result suggests that the hydrate/liquid transition is much faster when confined in porous rocks, which is consistent with previous experimental data~\cite{linga_enhanced_2012, phan_aqueous_2014}. For bulk hydrate, in agreement with previous conclusion by Molinero and coworkers~\cite{knott_homogeneous_2012}, the very large free energy barrier at moderate $T - T_{f,0}$ leads to non-physical nucleation times $\tau$ beyond the age of the Universe. As noted in Ref.~\cite{knott_homogeneous_2012}, this result suggests that the formation of bulk methane hydrate observed in real conditions never occurs through homogeneous nucleation. This interpretation is supported by our data which point to a smaller $\Delta \Omega^\ast$ for surface nucleation under the same thermodynamic conditions.

To provide a quantitative estimate of the formation/dissociation kinetics for confined hydrate, let us consider the experimental situation of a porous medium filled with liquid water and gas methane in equilibrium with a bulk reservoir. We assume that the system is brought to temperature $T$ within a time insufficient to allow hydrate formation within the bulk phase (justified as typical experiments are performed on much shorter timescales than the characteristic nucleation time for bulk hydrate). With such conditions, both the confined and bulk phases are metastable when reaching $T < T_f(D_p)$.  Using the nucleation theory, the nucleation rate $1/\tau = J \times A$  for surface and confined nucleation can be expressed as the surface area $A$ developed by the porous material multiplied by the number of nuclei formed per unit of surface and unit of time, $J= \rho_s D Z \exp\left[-\Delta \Omega^\ast/k_BT\right]$  ($\rho_s$ is the nucleus surface density, $D_s$ the diffusivity, and $Z$ the Zeldovich factor which accounts for the probability that a nucleus at the top of the free energy barrier does not melt and actually ends up forming hydrate). Assuming the same prefactor $\rho_s D Z$ for surface and confined nucleation, the latter expression allows comparing the  nucleation times for these two mechanisms from the porous surface area $A_p$ and external surface area $A_e$. For weak metastability [small $T_f(D_p) - T$], the confined nucleation time $\tau_c$ is much faster than at the external surface $\tau_s$ because $\Delta \Omega_c^\ast < \Delta \Omega_s^\ast$ and $A_p >> A_e$. Similarly, for strong metastability [large $T_f(D_p) - T$], even if $\Delta \Omega_c^\ast = \Delta \Omega_s^\ast$, $\tau_c << \tau_s$ because $A_p >> A_e$. Regardless of the exact confined nucleation mechanisms (hemispherical cap \textit{versus} bridge), considering that $A_p$ is proportional to the pore size to volume ratio and, hence, to the reciprocal pore size, the nucleation time is expected to scale as $\tau_c \sim 1/D_p$. In particular, this simple model predicts that the ratio between the nucleation times for confined and non-confined methane hydrate is proportional to the ratio of the external to pore specific surface area, $\tau_c/\tau_s \propto A_e/A_p$. Despite its simplicity, this semi-quantitative picture is consistent with experimental observations showing that confinement reduces the formation time from weeks to hours~\cite{borchardt_methane_2018, casco_experimental_2019}. These calculations depend on the exact choice made for $\theta$ but our conclusions remain valid for all angles as shown in  Fig. S9(b).

\section*{Discussion}

A molecular simulation approach is used to show that the thermodynamics of methane hydrate confined down to the nanoscale conforms the classical macroscopic picture as described by the Gibbs--Thomson equation. When confined in pores of a few molecular sizes, even if water and methane possess different wetting properties towards the host surface, they form a gas hydrate phase with structural properties very close to their bulk counterpart. A negative shift in the melting point is observed as a result of the pore/liquid surface tension being smaller than the pore/hydrate surface tension. Beyond thermodynamic aspects, by showing that confinement leads to facilitated methane hydrate formation and dissociation, our free energy calculations and mesoscopic nucleation model provide a theoretical support for the faster kinetics observed in experimental studies. More importantly, by relying on a statistical physics approach at the molecular level, our study provides microscopic insights into the formation/dissociation nucleus and the key role of the host porous surface.

Together with the large body of experimental data available, the present results provide a unifying picture of methane hydrate in confined environments or in the vicinity of surfaces. While the description of confined methane hydrate adopted here is simplified compared to real systems, our findings remain meaningful to physical situations such as methane hydrate trapped in rocks on Earth or in other planets, comets, etc. They are also relevant to important energy and environmental aspects such as the expected impact of methane hydrate dissociation on global warming, the use of methane hydrate as energy storage devices, the formation of methane hydrate in pipelines, etc. Despite the simple approach used in this work, in addition to the melting point depression being consistent with available experimental data, the predicted methane hydrate formation and dissociation times provide a robust molecular scale picture of experiments in this field. In particular, considering the mechanisms and times associated with homogeneous and heterogeneous nucleation, methane hydrate is expected to form in the vicinity of pore surface and then to extend to the bulk external phase. In this context, beyond the temperature-dependent crossover between confined and surface nucleation, the predicted scaling between the induction time $\tau \sim 1/D_p$ and pore size $D_p$ is an important result which sheds light on complex experimental behavior.

Despite these results, there is a number of aspects which were not considered explicitly in our approach. As shown by Borchardt and coworkers~\cite{casco_experimental_2019}, even when very similar porous samples are considered, the detailed surface chemistry is expected to play a key role on the thermodynamics and kinetics of methane hydrate formation in confinements. In particular, while our approach only considered the formation (dissociation) of methane hydrate  from (towards) liquid water, these authors observed using advanced structural characterization tools more complicated mechanisms involving the formation and distribution of hexagonal and/or disordered ice within the sample porosity. Moreover, while diffusion limitations of water and methane through the porosity is expected to affect the kinetics of methane hydrate formation and dissociation~\cite{zhao_growth_2020}, they are not taken into account in our statistical mechanics approach which only considers thermodynamic aspects since mass transport and diffusion are not explicitly treated. However, in the classical nucleation theory, such diffusion limitations only affect the prefactor and the typical nucleation time remains mostly driven -- as a first order approximation -- by the free energy metastability barrier. More importantly, despite the drawbacks identified above, the fact that our molecular simulation approach provides a picture consistent with experiments in almost every point makes our approach suitable to tackle issues related to methane hydrate in confined geometries.

\matmethods{
\subsection*{Models} \label{subsec:MolecularModelling}

Methane is described as a single Lennard-Jones (LJ) site with parameters from the OPLS-UA force field~\cite{jorgensen_development_1996}. Water is described using the TIP4P/Ice rigid water model~\cite{abascal_potential_2005} which contains 4 sites: an LJ site on the oxygen, two point charges on the hydrogen atoms, and a site M which corresponds to the negative charge of the oxygen located at a distance $d_\textrm{OM}=$ 0.1577 {\AA{}} from the oxygen toward the hydrogen atoms along the H--O--H angle bisector. In this model, the O--H bond length is 0.9572 $\text{\AA{}}$ and the H--O--H angle 104.52$^\circ$. As shown in Refs.~\cite{conde_determining_2010, jin_molecular_2017}, TIP4P/Ice water combined with OPLS-UA methane accurately reproduces the experimental phase diagram of bulk methane hydrate. We use the stochastic procedure by Buch et al.~\cite{buch_simulations_1998} to generate a hydrate molecular configuration with an sI structure~\cite{chakraborty_amonte_2012, buch_simulations_1998, jin_molecular_2017}. The slit pores considered in this study are built as follows. Each pore surface is of lateral dimensions $L_x = L_y \sim 2.38$ nm and made up of $11 \times 11$ squares whose vertices and centers are occupied by a solid atom. Solid atoms are maintained frozen in all molecular simulations. The potential energy includes intermolecular interactions  but no intramolecular interactions as we consider a rigid water model, frozen solid, and united-atom methane model. The intermolecular potential between two atoms $i$ and $j$ includes a short-range repulsion and attractive dispersion described using the LJ potential: $u_{ij}^{LJ}(r) = 4\varepsilon_{ij} [(\sigma_{ij}/r)^{12} - (\sigma_{ij}/r)^6]$ where $r$ is the distance separating $i$ and $j$ while $\varepsilon_{ij}$ and $\sigma_{ij}$ are the corresponding parameters. The LJ interactions are truncated beyond $r_c = 1.19$ nm. The like-atom parameters are given in Table S1 while the unlike-atom parameters are obtained using the Lorentz--Berthelot mixing rules. The intermolecular potential between $i$ and $j$ also includes the Coulomb contribution: $u_{ij}^{C}(r) = q_i q_j/(4\pi \varepsilon_0 r)$ where $q_i$ and $q_j$ denote the atomic charges (Table S1). Ewald summation is used to correct for the finite system size with an accuracy $10^{-5}$.

\subsection*{Molecular Dynamics and Monte Carlo simulations}

Molecular Dynamics (MD) in the NPT ensemble is used to determine the molar volumes ($v^\textrm{L}$, $v^\textrm{H}$) and  enthalpies ($h_w^\textrm{L}$, $h_m^\textrm{V}$, $h_{m,w}^\textrm{H}$) at bulk coexistence conditions ($T_{f,0}$, $P_0$). The velocity-Verlet algorithm is used to integrate the equation of motion with an integration timestep of 1 fs. The temperature and pressure are controlled using Nos\'e-Hoover thermostat and barostat with a typical relaxation time of 2 ps for both NVT and NPT ensembles. All MD simulations were performed using \textit{Lammps}~\cite{plimpton_fast_1995}. Monte Carlo simulations in the grand canonical ensemble (GCMC) are used in our DCM to determine the confined transition temperature $T_f$. In this ensemble, the system has a constant volume $V$, methane and water chemical potentials $\mu_m$ and $\mu_w$, and temperature $T$. Monte Carlo moves in the grand canonical ensemble include molecule rotations, translations, insertions and deletions for both water and methane. A move from an old ($o$) to a new ($n$) microscopic states is accepted or rejected using a Metropolis scheme with an acceptance probability $P_{acc} = \min\{1, p_{\mu_m \mu_w V T}^n / p_{\mu_m \mu_w V T}^o \}$ where $p_{\mu_m \mu_w V T}$ corresponds to the density of states in the grand canonical ensemble: 
    \begin{equation}
        \begin{split}
        p_{\mu_m \mu_w V T}(\mathbf{s}^N) \propto & \frac{V^N}{\Lambda_m^{3N_m}\Lambda_w^{3N_w}N_m!N_w!} \\
        & \prod_{i=m,w}\exp\left(\frac{N_i\mu_i}{k_BT}\right) \exp \left(\frac{-U(\mathbf{s}^N)}{k_BT}\right)
        \end{split}
  \end{equation}
with $\Lambda_i=h/\sqrt{2\pi m_ik_BT}$ the thermal wavelength for molecule $i$ ($m$, $w$). $N_w$ and $N_m$ are the numbers of water and methane molecules, $\mathbf{s}^N$ is the coordinate set of the $N=N_w+N_m$ molecules in a microscopic configuration whose intermolecular energy is  $U(\mathbf{s}^N)$.

\subsection*{Free energy calculations}

Umbrella sampling is used to determine the free energy $\Omega$ as a function of the local order parameter $Q_6$. This quantity, which allows identifying liquid water and methane hydrate~\cite{radhakrishnan_new_2002}, is determined for a given oxygen O$_i$ as:
\begin{equation} \label{eq:DefinationParameter}
        Q_{6,i} = \left( \frac{4\pi}{13} \sum_{m=-6}^{6} \mid \mathbf{Q}_{6m,i} \mid^2 \right)^{1/2}
\end{equation}
where $\mathbf{Q}_{6m,i}$ are complex vectors defined as $Q_{6m,i} = 1/N_{b,i} \sum \mathbf{Y}_{6m}(\mathbf{r}_{ij})$. The summation over $j=1$ to $N_{b,i}$ runs for all neighbor oxygen within a distance $r_c^\prime=0.35$ nm) while $\mathbf{Y}_{6m} (\textbf{r}_{ij})$ are the spherical harmonics that depend on the vector $\mathbf{r}_{ij}=\mathbf{r}_j-\mathbf{r}_i$.
        
The free energy was determined using umbrella sampling with biased grand canonical Monte Carlo simulations. Starting from methane hydrate, we force its transformation into liquid water using a $Q_6$ dependent biasing potential $U^{B}(Q_6)$. Such biased simulations yield the biased probability distribution $P^B(Q_6)$ which can be corrected from the biased potential energy to determine the free energy as described in the main text. To sample the entire domain $Q$ (0.300--0.6), we consider $N_{s}=61$ windows with a spacing of 0.05 so that $N_{s}$ biased Monte Carlo simulations are carried out with the corresponding $Q_{6,k}^{0}$. In more detail, for the $k$-th window, the following biasing harmonic potential is used: $U^{B}_k(Q_6) = 1/2 K[Q_6-Q_{6,k}^{0}]^2$ where $K = 5\times10^7$ K is the force constant and $Q_{6,k}^{0}$ the central value. We use the weighted average of the unbiased probability distribution of each window $P_{k}^U(Q_6)$ to determine the full-unbiased probability distribution $P^U(Q_6)$:   
    \begin{equation}    \label{eq:ProDisUmbSam}
        P^U(Q_6) = \sum_{k=1}^{N_s} N_k P_k^U(Q_6) \exp\left(-\frac{U_k^B(Q_6)-\Omega_k(Q_{6})}{k_BT}\right)
    \end{equation}
where $P_k^U(Q_6)$ and $N_i$ are the unbiased probability distribution and the number of samples, respectively, in the $k$-th window. $\Omega_k(Q_6)$ is the metastability free energy which can be calculated according to:    
    \begin{equation}    \label{eq:FreEngUmbSam}
        \exp\left(-\frac{\Omega_k(Q_6)}{k_BT}\right) = \int dQ_6 P^U(Q_6) \exp\left(-\frac{U^B_k(Q_6)}{k_BT}\right)
    \end{equation}
Starting from \eqref{eq:ProDisUmbSam} with $\Omega_k(Q_6)=0$, we iterate self-consistently between Eqs.~(\ref{eq:FreEngUmbSam}) and (\ref{eq:ProDisUmbSam}) until convergence is reached.

} % END OF METHODS

\showmatmethods{} % Display the Materials and Methods section

\acknow{D. Jin acknowledges financial support from China Scholarship Council (CSC 201506450015). We thank F. Calvo, J. L. Barrat, C. Picard, L. Scalfi, and B. Rotenberg, L. Borchardt and S. Gratz, for interesting discussions.}

\showacknow{}

\end{document}